\documentclass{ifacconf}

\usepackage{natbib} 
\usepackage{amsmath}
\usepackage{graphicx}
\usepackage{amssymb}
\usepackage{cases}
\usepackage{subcaption}
\usepackage{empheq}
\usepackage{url}
\usepackage{subfiles}

\newcommand\RR{\mathbb{R}}

\newcommand\StateConstraintSet{\mathbb{X}}
\newcommand\InputConstraintSet{\mathbb{U}}

\newtheorem{remark}{Remark}

\begin{document}

\begin{frontmatter}
\title{Agentic MPC \\for Semantic Control System Resynthesis\thanksref{footnoteinfo}}
\thanks[footnoteinfo]{This work was supported by Grant-in-Aid for Scientific Research (B), No.~25K01254 from The Japan Society for the Promotion of Science (JSPS).}
\author{Yuya Miyaoka} 
\author{Masaki Inoue}

\begin{abstract} 
While MPC effectively handles structured, diverse, and low-level specifications, it lacks the capability to dynamically incorporate high-level contextual information such as social norms, user intent, or natural language instructions. To address this limitation, this manuscript introduces an agentic MPC framework that enables context-aware, semantically adaptive control synthesis by integrating with large language model-based agents. 
The agent interprets heterogeneous inputs, including natural language messages, environmental observations, and external knowledge, to resynthesize the control specifications. 
The effectiveness of the framework is demonstrated in an autonomous driving scenario, where the system aligns with personal preferences or responds to social situations such as emergency vehicle yielding. 
\end{abstract}
\end{frontmatter}

\section{Introduction}

Model Predictive Control (MPC) has demonstrated strong performance for several control applications by explicitly handling system dynamics, control objectives, and constraints to ensure safe, optimal operation.

Real-world driving not only optimizes control performance but also requires adaptive, context-aware decision-making in rapidly changing and ambiguous situations.
In practice, autonomous driving systems must continuously adjust their behavior depending on high-level contextual factors such as traffic conditions, surrounding objects, implicit social rules, and passenger preferences. 

Several studies have addressed these challenges by incorporating social attributes and human-like reasoning into autonomous driving systems.
In autonomous driving, works by \cite{Shi25_SocialAD, Zhao24_SocialAD, Crosato23_SocialAD, Hang21_SocialAD} introduce socially-aware and human-like decision-making mechanisms. 
In robotics, socially interactive systems are developed to generate human-like actions or explain their behavior (\cite{Valls25_SocialRobot, Siyao24_Duolando_SocialRobot, Stange22_SocialRobot}).
These works aim to generate behavior that is not only safe and efficient but also socially acceptable and interpretable.
However, these studies are limited in that the information used for decision-making is confined to a single modality, such as purely visual information or only the relative positions of surrounding vehicles and pedestrians. 
To achieve high-level contextual understanding, a unified mechanism capable of integrating heterogeneous sources of information, such as environmental perceptions, social contexts, and explicit human inputs, is required.

Among various information modalities, natural language has recently been engaged as an interface to incorporate such context and human intent into control systems.
For instance, recent studies explore the integration of natural language messages into control systems, enabling users to update control specifications. Specifically, works by \cite{Miyaoka26_ChatMPC, Wu25_InstructMPC, Sha25_LanguageMPC, Miyaoka24_ChatMPC} incorporate natural language processing (NLP) technology to update control specifications based on user-provided messages. 
Furthermore, some approaches focus on translating natural language requirements into formal languages, such as Signal Temporal Logic (\cite{Fushimi26_NL2STL, Fang25_NL2STL, Wang24_NL2STL}), allowing structured incorporation of semantic constraints into control systems.

To fully leverage these diverse information channels and semantic requirements, recent advances in AI agents, particularly those powered by large language models (LLMs), offer a promising breakthrough. These agents have shown remarkable capability in information integration and problem-solving within complex scenarios (see \cite{Wang25_AIAgent} and the references therein). The systems can interpret user inputs, retrieve external knowledge, and make context-sensitive decisions. 

This manuscript proposes a concept of \textit{agentic MPC} that interacts with an MPC controller to resynthesize control specifications. The AI agent observes user inputs and the current control specifications to understand the current context in which the controller operates. The agent is also allowed to modify or overwrite some specifications handled by the MPC controller. The resulting system combines the semantic adaptability of agents with the optimal control capability of MPC.
This manuscript also implements the agent that communicates with various information channels via the Model Context Protocol (MCP) introduced by \cite{Anthropic24_MCP} and validates the proposed framework in autonomous driving scenarios using the CARLA simulation developed by \cite{Carla}.

\section{Overview of Agentic MPC}\label{Section:AMPC}
To enable semantic context-aware adaptability in MPC, a promising approach is an agent.

\begin{figure}
    \centering
    \includegraphics[width=0.98\linewidth]{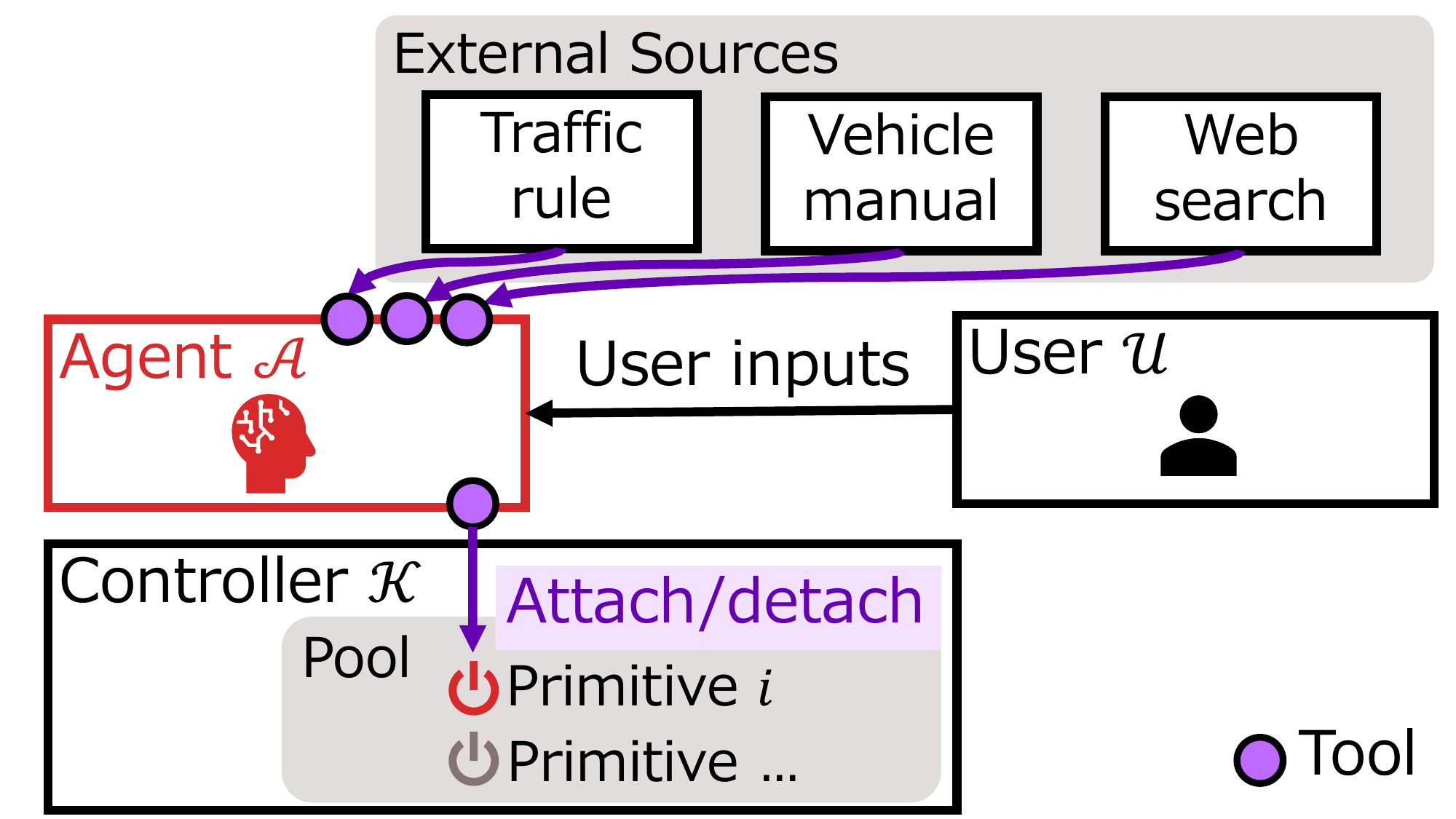}
    \caption{Concept of Agentic MPC.}
    \label{F:AMPC.Concept}
\end{figure}

As illustrated in Fig.~\ref{F:AMPC.Concept}, the agent operates alongside the MPC controller and continuously monitors the system context. The agent receives heterogeneous information sources, including user inputs, e.g., ``An ambulance is approaching from behind'', external observations, e.g., surrounding vehicles and obstacles, and optionally external knowledge sources, e.g., traffic rules, vehicle manuals, and web sources.
Based on such information, the agent interprets the current situation and determines whether modifications to the control specification are required.

To support flexible modification of the control specifications, the MPC controller is designed in a modular manner. Inspired by MPC Builder, developed by \cite{Honda23_MPCBuilder}, we decompose the control specifications into a set of reusable \textit{primitives}. Each primitive represents a minimal unit of objectives, constraints, or inner model. The agent selects some primitives for the controller to attach or detach. A predefined set of primitives is typically maintained, with all such primitives collectively referred to as the \textit{pool}.
Once the control specification is set, the MPC controller solves the resulting optimization problem using a standard solver. 

The connection between the agent and the MPC controller or some information channels is performed using invokable \textit{tools}. 
The tools have the following characteristics: 
\begin{itemize}
    \item They are invoked through discrete commands,
    \item Each command can accept quantitative arguments, e.g., reference speed,
    \item Invoking a tool results in defined actions, e.g., retrieving thevehicle manual or modifying control specification.
\end{itemize}
Some tools correspond to operations that attach or detach primitives to/from the controller. This design achieves limited flexibility: while the accessible tools are constrained for safety, their quantitative arguments allow continuous adaptation to a variety of scenarios.

\section{General Formulation of Agentic MPC}
This section presents the detailed design of the proposed framework, including the MPC formulation and the structure of the agent.

\subsection{Plant}
We consider the time-discrete state-space model:
\begin{align}
    x(k+1) =  f(x(k), u(k)), 
\end{align}
where $x(k) \in \RR^N$ is the current state, $u(k) \in \RR^M$ is the control input, and $f: \RR^N \times \RR^M \to \RR^N$ represents the plant dynamics. 

\subsection{MPC Formulation}
\subsubsection{Basic Formulation.}
Given a control specification defined by the primitives, the MPC solves an optimization problem to obtain the optimal control input sequence. The optimization problem at time $t$ is formulated as:

\begin{subequations}
\begin{empheq}[left={\empheqlbrace}]{alignat=1}
    \min_{ \hat u(t) } ~& J(\hat u(t)) , \\
    \text{s.t.} ~
        & \hat x(0|t) = x(t), \\
        \begin{split} & \hat x(k+1|t) = \hat x(k|t) + f(\hat x(k|t), \hat u(k|t)), \\ 
            & \quad \forall k \in\{0,\ldots,H-1\} \end{split}\\
        & \hat x(k|t) \in \StateConstraintSet, 
            \ \forall k \in \{1,\ldots,H\}, \\
        & \hat u(k|t) \in \InputConstraintSet,
            \ \forall k \in \{0,\ldots,H-1\} , 
\end{empheq}\label{E:Design.MPCOriginalProblem}
\end{subequations}

where $H$ is the prediction horizon, $\hat x(k|t) \in \RR^N$ is the predicted state, $\hat u(k|t) \in \RR^M$ is the predicted control input, and $\hat u(t):=[\hat u(0|t) ~ \cdots ~ \hat u(H-1|t)]$ is the input sequence to be determined in this problem.
The control specifications are represented by objective function $J : \RR^{H\times M} \to \RR$, state constraint $\StateConstraintSet \subseteq \RR^N$, and input constraint $\InputConstraintSet \subseteq \RR^M$.

\begin{remark}
    The objective function $J$ is an abstract formulation, and is generally implemented by dividing it into stage cost $J_\text{stage}$ and final cost $J_\text{final}$:
    \begin{align}\begin{split}
        & J(\hat u(t)) \\
        & = \sum_{k=0}^{H-1} \left[ J_\text{stage}(\hat x(k|t), \hat u(k|t))  \right] 
        + J_\text{final}(\hat x(H|t)).
\end{split}\end{align}
\end{remark}

\subsubsection{MPPI, a Fast MPC Solver.}
The original MPC optimization problem \eqref{E:Design.MPCOriginalProblem} is fundamentally complex, with computational cost being a major concern. To address this issue, we employ Model Predictive Path Integral (MPPI, \cite{Asmar23_MPPI, Graby17_MPPI}). MPPI features a sampling-based optimization method for a predefined optimization problem and is characterized by high-speed operation via a parallelizable algorithm.

MPPI generates $S$ candidate control input sequences $\hat u_s(t), s\in\{1,\ldots,S\}$ by sampling from a Gaussian distribution $\mathcal N(\bar u, \sigma^2 I_H)$, where $\bar u$ is the mean and $\sigma^2\ge 0$ is the variance. The optimal control input sequence $u^\ast(t)$ is approximated as follows:
\begin{align}
    u^\ast(t) = \frac{\sum_{s=1}^S w_s \hat u_s(t) }{\sum_{s=1}^S w_s},
\end{align}
where $w_s$ is the weight and is given by 
\begin{align}
    w_s = \exp(- J(\hat u_s(t) / T)).
\end{align}

\begin{remark} 
 In MPPI, handling hard constraints explicitly within the optimization process is generally difficult due to its sampling-based algorithm. Therefore, state constraint $\StateConstraintSet$ and input constraint $\InputConstraintSet$ are incorporated into the objective function as soft constraints.
 Specifically, let $c_\text{state} : \RR^N \to \RR$ and $c_\text{input} : \RR^M \to \RR$ be state and input constraint violations, respectively. Then, the objective function used in MPPI $J_\text{MPPI}$ is given by:
 \begin{align}\begin{split}
     &J_\text{MPPI}(\hat u(t)) \\
     &=  J(\hat u(t)) + \sum_{k=1}^{H} c_\text{state}(\hat x(k|t)) + \sum_{k=0}^{H-1} c_\text{input}(\hat u(k|t)).
 \end{split}\end{align}
\end{remark}
\begin{remark} 
 MPPI does not solve the original control problem \eqref{E:Design.MPCOriginalProblem} exactly. MPPI is a sampling-based method, which introduces an implicit regularization term on the decision variables $\hat u(t)$. Thus, the resulting solution $u^\ast(t)$ tends to stay close to the sampling distribution mean $\bar u$, and large deviations are penalized. 
\end{remark}

\subsubsection{Extension of MPPI, Fourier MPPI.}
While MPPI provides strong performance, it suffers from the curse of dimensionality when the prediction horizon $H$ is long. To address this, we propose an extension called Fourier MPPI, which aims to reduce the sampling dimension.

 Instead of directly sampling the control sequence $\hat u_s(t) \in \RR^{H\times M}$, Fourier MPPI renders control inputs using a finite Fourier series:

 \begin{align}\begin{split}
     & u(k|t) = \beta_0 + \sum_{l=1}^L  
     \alpha_l \sin \left( \frac{2\pi l}{H} k \right) 
     + \beta_l \cos  \left( \frac{2\pi l}{H} k \right) , \\
     & \quad \forall k \in \{0,\ldots,H-1\} ,
 \end{split}\end{align}
 where $L$ is the truncation order, $\beta_0 \in \RR^M$, $\alpha_l \in \RR^M$, and $\beta_l \in \RR^M$, $l\in\{1,\ldots,L\}$ are Fourier coefficients to be sampled. Letting $z := [\beta_0~ \alpha_1~ \beta_1 ~\cdots~ \alpha_L ~\beta_L]^\top$, the optimal Fourier series of the control input sequence $z^\ast$ is given by:
 \begin{align}
    z^\ast = \frac{\sum_{s=1}^S w_s \hat z_s}{\sum_{s=1}^S w_s},
\end{align}
where $w_s$  is the weight calculated by 
\begin{align}
    w_s = \exp(- J_\text{MPPI} (\hat u_s(t) / T) )
\end{align}

 \subsubsection{Primitives of Control Specifications.}
The objective function used in the Fourier MPPI consists of small objective functions for each control requirement, such as path tracking, collision avoidance, or adherence to implicit driving preferences. 
In this manuscript, the primitives introduced in Section~\ref{Section:AMPC} correspond to these functions. The total objective function applied in the MPC is constructed as a sum of multiple primitive objective functions.

Let $\mathcal J$ denote a set of all primitive objective functions available, let $\mathcal J_\text{attached}$ denote a set of attached objective functions that are included within the total objective function, and let $J_\text{MPPI,0}$ denote a mandatory objective function that cannot be detached. Then, the total objective function $J_\text{MPPI}$ is defined as:

\begin{align}
    J_\text{MPPI}(\hat u(t)) = J_\text{MPPI,0} + \sum_{J_\text{p} \in \mathcal J_\text{attached}} J_\text{p} (\hat u(t)) .
\end{align}

Note that all constraints are incorporated into the objective function as soft constraints, since the MPC controller internally employs Fourier MPPI.

\subsection{Agent}
The proposed framework employs an agent to enable semantic interpretation and adaptive modification of control specifications.
The agent is driven by a large language model (LLM), which provides strong semantic understanding and general knowledge to interpret diverse contextual information.

\subsubsection{States.}
The agent can be interpreted as a discrete-time dynamical system. 
Let $\tau \in \{0,1,\ldots\}$ denote the \textit{agent step}, which evolves independently from the control step $t$. At each agent step $\tau$, the agent maintains a message history $\mathcal H$:

\begin{align}
\mathcal H(\tau) = [m(0), m(1), \ldots, m(\tau)] ,
\end{align}
where each message $m(\tau)$ belongs to one of the following types:
\begin{description}
    \item[System message:] Predefined instructions or system-level context,
    \item[User message:] User inputs and externally provided information, 
    \item[Agent message:] Agent-generated responses, including text and tool invocation requests, 
    \item[Tool result:] [Optional] Outputs provided by invoked tools.
\end{description}
The history $\mathcal H(\tau)$ serves as the state of the agent, containing past observations and past decisions. Therefore, the agent can leverage previously processed context and its own past actions when making new decisions. 

\subsubsection{State Update.}
A new agent step is triggered by new messages. 
When either a new user message or a tool result $m$ is appended to the history at $\tau$, i.e., $\mathcal H(\tau + 1) = [\mathcal H(\tau), m]$, the agent then generates an agent message and appends it immediately: 
\begin{align}
    (m^\text{text}(\tau+2), m^\text{tool}(\tau+2)) = \mathcal A(\mathcal H(\tau+1)), 
\end{align}
where $\mathcal A$ is the agent, $m^\text{text}$ and $m^\text{tool}$ are the text and tool invocation request, respectively.

If tool invocations are issued by $m^\text{tool}$, corresponding tool result messages are generated and appended to the history, yielding more state updates. This process may iterate multiple times unless further tool invocations are required by the agent message. If no tool invocation is issued, the agent step turns idle until a new user message is received.

\subsubsection{Tools.}
Recalling the tool definition in Section~\ref{Section:AMPC}, the agent and the external channels, including the MPC controller, should interact via the tool.
To ensure structured interaction with the control system, the agent communicates through a set of predefined \textit{tools} using MCP. 
MCP is a standard syntax for listing and invoking tools, allowing the agent to interact with the external channels in a unified manner. 

Some tools enable access to external knowledge sources, such as web searches or document searches.
More importantly, several tools interact with the MPC controller; they attach or detach corresponding primitive objective functions within the MPC controller. The examples for the tools are provided in Section~\ref{Section:Experiment}.

\section{Implementation and Case Study of Agentic MPC}\label{Section:Experiment}
This section presents the case study of the proposed framework in autonomous driving scenarios to verify its effectiveness.
The experiments are conducted in the CARLA simulator developed by \cite{Carla} using the vehicle model corresponding to the blueprint-id ``vehicle.nissan.patrol''. 

\subsection{Control Objectives}
\begin{figure}
    \centering
    \includegraphics[width=1\linewidth]{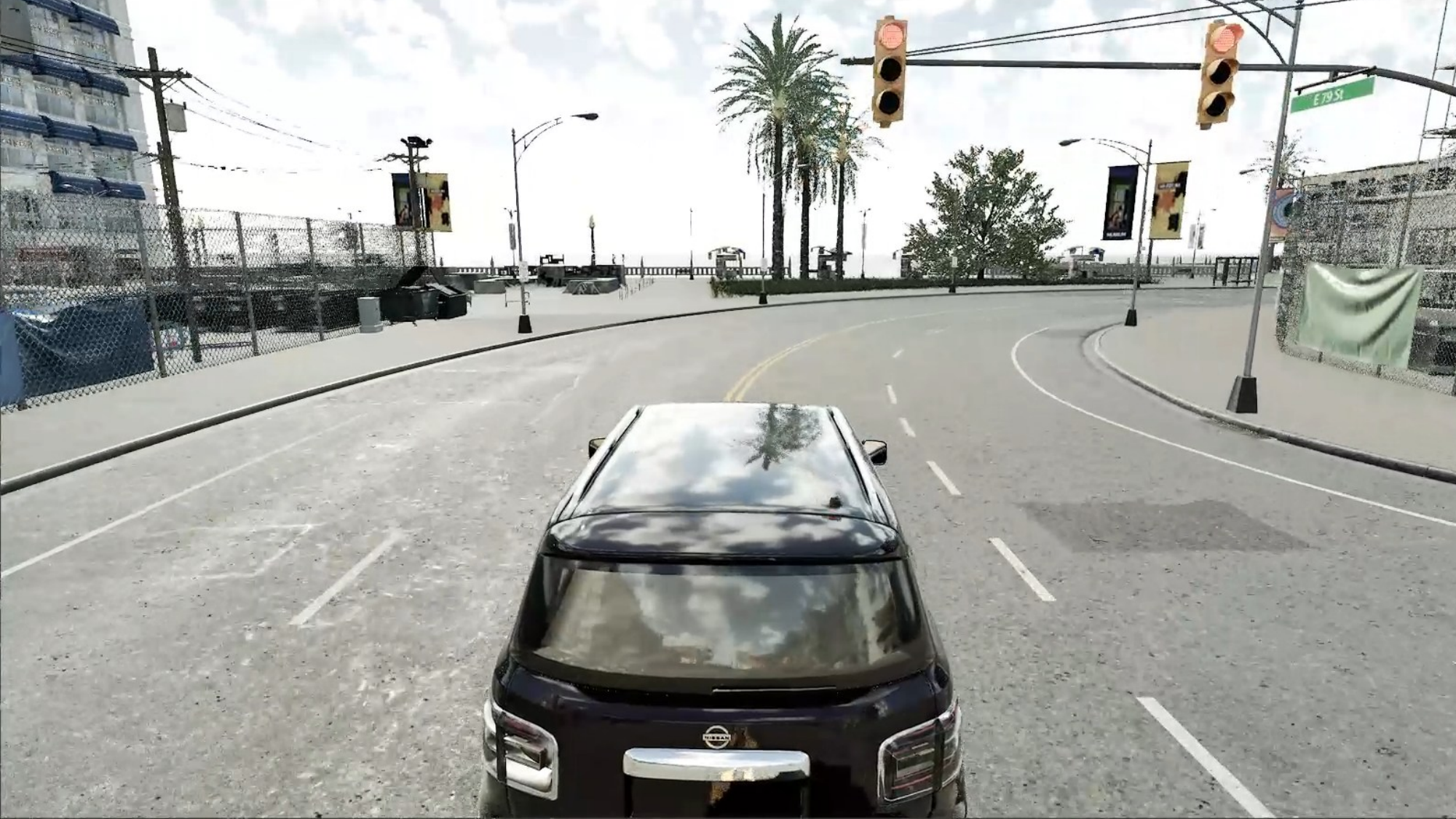}
    \caption{The simulation environment.}
    \label{F:Experiment.BirdView}
\end{figure}

The simulation environment is shown in Fig.~\ref{F:Experiment.BirdView}. The goal of the task is to navigate the vehicle along the center of the road from a designated start position to a designated goal position.

\subsection{MPC Setup}
\subsubsection{Vehicle Dynamics.}
The state of the vehicle is defined as
\begin{align}
    [x(k) ~ y(k) ~ \phi(k) ~ v(k)]^\top \in \RR^4,
\end{align}
where $(x,y)$ is the position, $\phi$ is the heading angle, and $v \ge 0$ is the speed. The control input is given by
\begin{align}
    u(k) = [\delta(k) ~ a(k) ~ b(k)]^\top \in \RR^3,
\end{align}
where  $\delta$ is the steering input, $a \in [0,1]$ is the throttle, and $b \in [0,1]$ is the brake input. The sampling interval is $\Delta t = 0.04~\text{s}$.

The vehicle dynamics are composed of two parts: a kinematic bicycle model (KBM) presented by \cite{Philip17_KBM}, which describes the vehicle's position and heading angle dynamics, and a simple linear model that describes the vehicle's speed dynamics. The vehicle dynamics are formulated as:
\begin{subequations}
\begin{empheq}[left={\empheqlbrace}]{alignat=1}
x(k+1) &= x(k) + \Delta t \ v(k) \cos(\phi(k) + \gamma(k)), \\
y(k+1) &= y(k) + \Delta t \ v(k) \sin(\phi(k) + \gamma(k)), \\
\phi(k+1) &= \phi(k) + \Delta t \ \frac{v(k)}{l_r} \sin(\gamma(k)), \\
\begin{split}
v(k+1) &= \max [ 0, \\
& v(k) + \Delta t \left( w_\text{a} a(k) + w_\text{b} b(k) + w_\text{v} v(k) \right) ] ,
\end{split}
\end{empheq}\label{E:Experiment.Dynamics}
\end{subequations}
where $\gamma(k)$ is the slip angle at the center of the vehicle and given by:
\begin{align}
    \gamma(k) = \arctan \left( \frac{l_r}{l_f+l_r} \tan(r_\delta \delta(k)) \right) .
\end{align}

The model parameters are identified as follows:
\begin{align}\begin{array}{llll}
    & r_\delta = 0.1306,  & l_r = 8.184, & l_f = -4.114, \\
    & w_\text{a} = 4.615, & w_\text{b} = -5.980, & w_\text{v} = -0.07220 .
\end{array}\end{align}

\subsubsection{Fourier MPPI.}
The solution to the perturbed MPC problem \eqref{E:Design.MPCOriginalProblem} is obtained using Fourier MPPI.
In Fourier MPPI, two Fourier coefficient sequences are sampled: one is $z_\delta$, which generates the steering input sequence $\{\delta(k)\}_{k=0}^{H-1}$, and another one is $z_\text{ab}$, which generates both the throttle input sequence $\{a(k)\}_{k=0}^{H-1}$ and the brake input sequence $\{b(k)\}_{k=0}^{H-1}$. 
We set the truncation order of the Fourier series as  $L = 4$, which means that the Fourier coefficient sequence is a vector of order 9. 
The sampling variance is scaled depending on the coefficient order. The standard deviations for each coefficient are set as:
\begin{align}\begin{array}{llll}
 & \beta_{\delta, 0} : 0.8, & \alpha_{\delta, l} , \beta_{\delta, l} : 0.8 \cdot 0.8^l ,  & \forall l \in \{1,\ldots,L\} , \\
 & \beta_{\text{ab}, 0} : 0.3, & \alpha_{\text{ab}, l} , \beta_{\text{ab}, l} : 0.3 \cdot 0.8^l , & \forall l \in \{1,\ldots,L\} .
\end{array}
\label{E:Experiment.SampleSD}
\end{align}
The means $\bar \beta_\delta$ and $\bar \beta_\text{ab}$ are set to the optimal Fourier coefficients that are calculated in the previous step.
The number of samples is set to $S  = 256$.

Given the Fourier coefficient sequences, the control input sequence is constructed as follows:
\begin{subequations}
\begin{empheq}[left={\empheqlbrace}]{alignat=1}
    \begin{split}
        \delta(k) &= \beta_{\delta, 0} \\ &+ \sum_{l=1}^L 
        \alpha_{\delta, l} \sin \left( \frac{2\pi l}{H} k \right) + \beta_{\delta, l} \cos  \left( \frac{2\pi l}{H} k \right) ,
    \end{split} \\
    \begin{split}
        a(k) &= \text{Sat}_{[0,1]} [   \beta_{\text{ab},0} \\
            & + \sum_{l=1}^L
            \alpha_{\text{ab}, l} \sin \left( \frac{2\pi l}{H} k \right) 
            + \beta_{\text{ab}, l} \cos  \left( \frac{2\pi l}{H} k \right)  \\
        & ], 
    \end{split}\\
    \begin{split}
    b(k) &= - \text{Sat}_{[-1,0]} [   \beta_{\text{ab},0} \\
        & + \sum_{l=1}^L 
        \alpha_{\text{ab}, l} \sin \left( \frac{2\pi l}{H} k \right) 
        + \beta_{\text{ab}, l} \cos  \left( \frac{2\pi l}{H} k \right) \\
    & ],
    \end{split}
\end{empheq}\label{E:Experiment.InputSequenceGeneration}
\end{subequations}
for all $k \in \{1,\ldots,H\}$. Here, $\text{Sat}_{[a,b]}[x]$ is the saturation function that constrains $x$ within $[a,b]$. 

\subsubsection{Primitive Objective Functions.}
The pool has three primitive objective functions, $J_1$, $J_2$, and $J_3$, as following:

\begin{description}
\item[Path Tracking Objective $J_1$:] The objective function is designed to make the vehicle follow a given path at a reference speed $v_\text{ref}$. In particular, we aim for the predicted trajectory to track a sequence of reference points:
\begin{align}
    J_1(\hat u) = \frac{1}{2} \sum_{k=1}^H 
    \left\|
        (\hat x(k), \hat y(k)) - (x_\text{ref}(k), y_\text{ref}(k)) 
    \right\|_2^2 ,
\end{align}
where $(x_\text{ref}(k), y_\text{ref}(k))$, $k \in \{1,\ldots,H\}$ is a sequence of reference points.

These reference points $(x_\text{ref}(k), y_\text{ref}(k))$ represent the motion of a point traveling along a given geometric path $(x_\text{path}(\tau), y_\text{path}(\tau))$ with the parameter $\tau \in [0,1]$ at a constant speed $v_\text{ref}$.

Formally, let $(x_\text{path}(\tau), y_\text{path}(\tau))$, $\tau \in [0,1]$, denote the geometric path, and let $\tau_0$ correspond to the closest point. The parameter $\tau_k$ which corresponds to the reference point $(x_\text{ref}(k), y_\text{ref}(k))$, satisfies:
\begin{align}
    \int_{\tau_0}^{\tau_k} \sqrt{ 
        \left( \frac{dx_\text{path}}{d\tau}(\tau')\right)^2 +  
        \left(\frac{dy_\text{path}}{d\tau}(\tau') \right)^2
    }  \, d\tau' 
    = v_\text{ref} \Delta t \, k.
\end{align}
This formulation ensures that the consecutive reference points $(x_\text{ref}(k), y_\text{ref}(k))$ and $(x_\text{ref}(k+1), y_\text{ref}(k+1))$ are separated by a distance of $v_\text{ref}\Delta t$.

\item[Obstacle Avoidance Objective $J_2$:] A penalty is applied when the predicted trajectory enters an obstacle region. Let $(x_\text{c}, y_{c})$ and $r$ be the center and radius of an obstacle region to be avoided. Then, the objective function is defined as:
\begin{align}
    J_2(\hat u) =  \frac{100}{H} \sum_{k=1}^H I(\hat x(k); x_\text{c}, y_\text{c}, r),
\end{align}
where $I$ is the indicator function that takes $1$ if the position $\hat x(k)$ enters the obstacle region and takes $0$ if not.

\item[Preferred Steering Direction Objective $J_3$:] This primitive objective function yields a soft bias encouraging steering in a given direction.
\begin{align}
    J_3(\hat u) = \begin{cases}
        \frac{1}{2H} \ \sum_{k=0}^{H-1} \min(0, \hat \delta(k))^2, & p=\text{left} , \\
        \frac{1}{2H} \ \sum_{k=0}^{H-1} \min(0, - \hat \delta(k))^2, & p=\text{right} ,
    \end{cases}
\end{align}
where $p \in\{\text{left}, \text{right}\}$ is the preference which direction to upvote.
\end{description}

\subsection{Agent Setup}
The agent is triggered by user messages and operates asynchronously with respect to the MPC control loop. The agent is provided (A) currently attached primitive objective functions and (B) available tools with the system prompt. 
The LLM applied in the agent is IBM Granite-4.1 8B \cite{IBM26_Granite41}.

\subsubsection{Available Tools.}
The agent interacts with the MPC controller via predefined tools:
\begin{description}
    \item[run\_center\_lane:] Attaches the path tracking objective function $J_1$. In the function, the reference path is set to the center lane. This tool takes one argument, the reference speed $v_\text{ref}$, which is used in the objective function. This tool also detaches the other path tracking objective function.
    \item[run\_left\_lane:] Attaches the path tracking objective function. In the function, the reference path is set to the left lane. The arguments and the other functions are the same as ``run\_center\_lane''.
    \item[set\_preferred\_steering\_direction:] Attaches the preferred steering direction objective function. This tool takes one argument, the preferred direction $p \in\{\text{left}, \text{right}\}$, which is used in the objective function.
\end{description}

\subsection{Scenario 1 - Steering Preference}
This scenario demonstrates an offline modification that incorporates the passenger's preference.

A circular obstacle is placed on the reference path. The initial attached primitives include the path tracking objective function $J_1$, with the preference path being the center lane and the reference speed $v_\text{ref}=15~\text{m/s}$, and the obstacle avoidance objective function $J_2$. 
At the agent step \underline{$\tau=0$}, the system message is appended, and its content includes the initially attached primitive objective functions.

\begin{figure}[tbp]
  \centering
  \begin{minipage}{.48\linewidth}
    \centering
    \includegraphics[width=\linewidth]{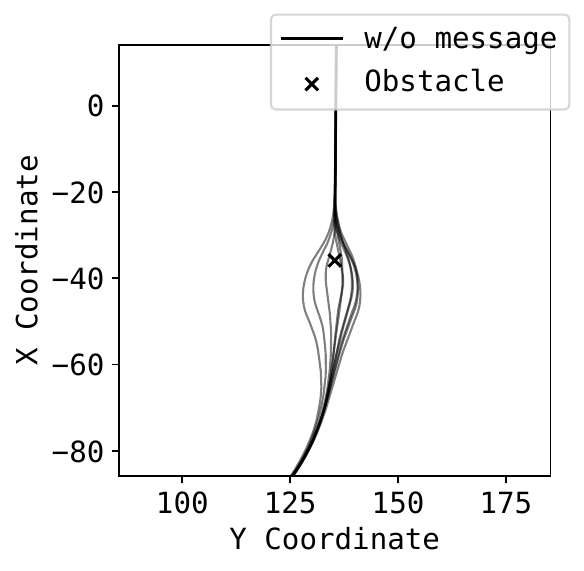}
    \subcaption{W/o passenger message}
    \label{F:Experiment.S1.Baseline}
  \end{minipage}
  \begin{minipage}{.48\linewidth}
    \centering
    \includegraphics[width=\linewidth]{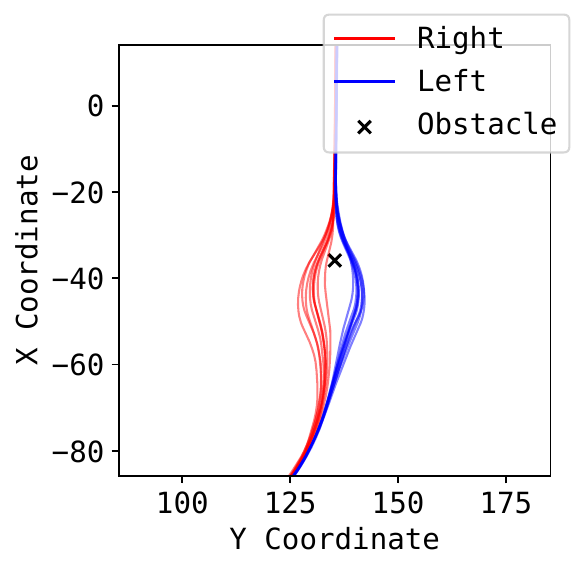}
    \subcaption{With passenger message}
    \label{F:Experiment.S1.WithMessage}
  \end{minipage}
  \caption{Resulting trajectories for Scenario 1}
  \label{F:Experiment.S1}
\end{figure}

First, ten trials are conducted without any user messages. In this case, the agent step remains at $\tau=0$. The result is shown in Fig.~\ref{F:Experiment.S1.Baseline}. In this figure, the vehicle is traveling from the top to the bottom. The vehicle passes the right side of the obstacle 3 times and the left side 7 times. This variability is caused by the stochastic nature of Fourier MPPI.

Next, two types of user messages are considered:
\begin{description}
    \item[User Message 1]  ``Let's try to avoid obstacles from the \underline{right}, shall we?''
    \item[User Message 2] ``Let's try to avoid obstacles from the \underline{left}, shall we?''
\end{description}
For each case, the user message is provided at the agent step \underline{$\tau=1$}. Consequently, at \underline{$\tau=2$}, the agent generates the message that invokes ``set\_preferred\_steering\_direction'' with the corresponding preferred direction. Specifically, for the User Message 1 case, it is set to ``right'', and for the User Message 2 case, it is set to ``left''. As a result, the preferred steering direction objective function $J_3$ is attached with the specified direction.

The resulting trajectories are shown in Fig.~\ref{F:Experiment.S1.WithMessage}. In this figure, the red lines indicate the User Message 1 case, and the blue lines indicate the User Message 2 case. In the User Message 1 case, the vehicle passes on the right in all trials. In the User Message 2 case, the vehicle passes on the left in all trials. 

After the user message, i.e., $\tau=2$, the agent successfully interprets the semantic meaning of the user message and translates it into an appropriate modification of the control specification.

\subsection{Scenario 2 - Emergency Response}
This scenario demonstrates an online modification that responds to a situation requiring only temporary changes to control specifications.

Assume that an ambulance is coming behind the operating vehicle. The initial attached primitive is the path tracking objective function $J_1$, with the preference path being the center lane and the reference speed $v_\text{ref}=15~\text{m/s}$.

The following two messages are provided at intervals: 
\begin{description}
    \item[First user message:] ``An ambulance is coming behind us. We'd better let them pass.''
    \item[Second user message]  ``Thank you. The ambulance is gone.''
\end{description}

\begin{figure}
    \centering
    \includegraphics[width=1\linewidth]{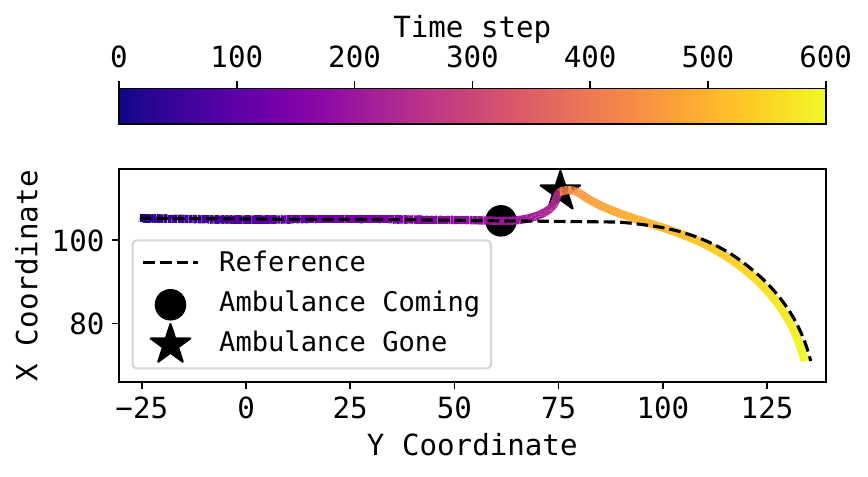}
    \caption{Resulting trajectory for Scenario 2}
    \label{F:Experiment.S2}
\end{figure}

Figure~\ref{F:Experiment.S2} illustrates the resulting trajectory. 
Initially, at \underline{$\tau=0$}, the system message is appended, and its content includes the initial attached primitive.
At \underline{$\tau=1$}, the user message ``An ambulance is coming behind us. We'd better let them pass.'' is received.
Consequently, at \underline{$\tau=2$}, the agent invokes the tool ``run\_left\_lane'' with the reference speed $v_\text{ref}=0$. As a result, the path tracking objective function $J_1$, with the preference path being the left lane and the reference speed $v_\text{ref}=0$, is attached. 
At \underline{$\tau=3$}, the second user message ``Thank you. The ambulance is gone.'' is provided.
As a result, at \underline{$\tau=4$}, the agent invokes the tool ``run\_center\_lane'' with the reference speed $v_\text{ref}=15~\text{m/s}$. As a result, the path tracking objective function $J_1$, with the preference path being the center lane and the reference speed $v_\text{ref}=15~\text{m/s}$, is attached. 

After the first message, i.e., $\tau=2$, the agent correctly interprets the semantic meaning of the situation, the approach of an emergency vehicle, and translates it into an appropriate objective function and its arguments. This demonstrates that the LLM-based agent can incorporate social rules into the control specifications.
After the second message, i.e., $\tau=4$, the system successfully restores the original control specification. This recovery is enabled in the system prompt, which includes the initially attached primitive objective functions. By retaining this information, the agent enables consistent and reversible adaptation.

\section{Conclusion}
This manuscript introduces an agentic MPC framework that bridges the gap between high-level semantic reasoning and low-level optimal control. By leveraging an LLM-based agent, the proposed system enables resynthesis of control specifications in response to heterogeneous information sources, including natural language inputs and contextual environmental cues.
A key contribution is the modularization of control specifications into primitives, allowing flexible and interpretable reconfiguration of control objectives. 
The autonomous driving scenarios are simulated to demonstrate the framework's ability to adapt behavior according to user preferences and implicit social rules, highlighting variability in real-world environments.

\section*{DECLARATION OF GENERATIVE AI AND AI-ASSISTED TECHNOLOGIES IN THE WRITING PROCESS}
During the preparation of this work, the authors used Gemini, TranslateGemma, and Plamo Translate in order to translate and refine the draft. After using this tool/service, the authors reviewed and edited the content as needed and take full responsibility for the content of the publication.

\bibliography{reference}

\end{document}